\documentclass{article}
\usepackage{graphicx} 
\usepackage[a4paper, top=30mm, bottom=30mm, left=30mm, right=30mm]{geometry}

\usepackage{color}
\usepackage{listings} 
\usepackage{ascmac} 
\usepackage{fancybox} 
\usepackage{amssymb} 
\usepackage{url} 
\usepackage{algorithm} 
\usepackage{algpseudocode} 
\usepackage{amsmath} 
\usepackage{multirow} 
\usepackage{bm} 
\usepackage[subrefformat=parens]{subcaption} 
\usepackage[colorlinks=true, linkcolor=blue, citecolor=blue, urlcolor=blue]{hyperref} 
\usepackage{enumitem}
\usepackage{booktabs} 
\usepackage{tikz} 
\usetikzlibrary{decorations.pathmorphing} 
\usepackage{enumitem} 
\usepackage{arydshln} 
\usepackage[square,numbers]{natbib} 

\title{Interoceptive Divergence in Aesthetic Evaluation\\ and Implications for Human-AI Alignment}
\author{Yoshia Abe, Tatsuya Daikoku, Yasuo Kuniyoshi}
\date{~~~}

\begin{document}

\maketitle

Artificial intelligence (AI), exemplified by large language models (LLMs), is rapidly approaching and in some cases surpassing human performance across a wide range of cognitive tasks. 
However, human nature is not limited to intelligence alone; 
it also encompasses sensibility, including the capacity to perceive and experience beauty in visual scenes or abstract ideas. 
This raises a fundamental question: how humans and AI systems converge or diverge in such aesthetic experiences.
As part of ongoing efforts in AI alignment aimed at harmonizing AI systems with human values, this study investigates the similarities and differences between humans and AI in aesthetic evaluation tasks involving visual stimuli. 
Aesthetic evaluation depends not only on objective properties of images but also on internal processes within the observer. 
Building upon prior human studies that have examined the relationship between beauty ratings, bodily sensations, and emotions, we adopt a comparable set of questionnaire items and present them to LLMs, enabling a direct comparison between human and AI responses. 
Our comparative analyses revealed that, while humans and AI exhibited broadly similar patterns in the correlations between beauty ratings and emotions, as well as in the image features they prioritized, notable divergences emerged in both the distribution of emotional responses and the relationship between beauty ratings and bodily sensations. 
These findings suggest that state-of-the-art LLMs, trained on large-scale textual data, can approximate average human tendencies in aesthetic evaluation to a certain extent. 
However, they also indicate limitations, particularly in relation to interoceptive aspects, which may reflect insufficient representation in training data or unintended consequences of alignment processes. 
These findings highlight key challenges for AI alignment and suggest important directions for developing AI systems with human-like aesthetic processing.

\section{Introduction}\label{sec1}

Artificial intelligence (AI), particularly large language models (LLMs), has approached--and in some cases surpassed--human-level performance across a wide range of domains (e.g., in higher-order cognitive tasks such as scientific reasoning and systematic thinking~\citep{latif2024openaio1outperformhumans}).
This development prompts a reconsideration of the human position in these domains.
These tasks nevertheless remain fundamentally intellectual and can be evaluated using objective, quantifiable metrics.
Human nature, by contrast, is not confined to intellectual abilities alone. 
In everyday life, humans perceive beauty and pleasantness in scenes and ideas they encounter. 
In tasks that require such sensibility, the relevant question is not whether humans or AI are superior, but how closely AI resembles humans. 
It is sometimes debated whether non-human animals may also possess similar sensibilities, but what about AI, which has been trained on data prepared by humans? 
To address this question, we compare the behavior of humans and AI, particularly LLMs, from the perspective of aesthetic evaluation of visual stimuli. 
Through this comparison, we aim to identify potential misalignment between human aesthetic values, shaped through the embodied developmental process, and those of AI, acquired through learning from textual data.

The concept of beauty has been extensively studied from the perspective of experimental aesthetics, including approaches in neuroscience and psychology~\citep{wassiliwizky2021why}.
Numerous studies have investigated human behavior in aesthetic evaluation of visual stimuli, including those examining patterns of aesthetic experience through behavioral experiments (as summarized in~\citep{palmer2013visual}), as well as studies that have identified core brain regions involved in the perception of beauty using neuroscientific methods~\citep{kawabata2004neural, ishizu2011toward}.
On the other hand, computational aesthetics has emerged as an approach to approximate the process of aesthetic evaluation using computational models~\citep{hoenig2005defining}, and with the advent of deep learning technologies, a variety of AI models have been proposed~\citep{valenzise2022advances}.
Within the era dominated by deep learning, studies by Iigaya et al.~\citep{iigaya2021aesthetic, iigaya2023neural} have examined the similarity between humans and AI in aesthetic evaluation of visual stimuli. 
Through comparisons between brain activity measurements and the processing of deep convolutional neural networks, their work has provided insights into the neural mechanisms underlying aesthetic value computation, such as the hierarchical representation of low- and high-level features and their integration in computing aesthetic value. 

In the 2020s, the dominant paradigm in AI models has shifted to LLMs. 
By learning from vast amounts of textual data, LLMs have acquired human-like general knowledge as well as capabilities for semantic and contextual understanding~\citep{hollmann2023large}, and have increasingly become capable of processing inputs from the image modality~\citep{yin2024mllm}.
Although LLMs produce outputs in the modality of text, through integration with external tools they are evolving into agents that can actively interact with and influence human society~\citep{xi2025rise}.
Accordingly, efforts in AI alignment, which aim to align the behavior of LLMs with human values, have become increasingly important~\citep{ji2025aialignment}.

Among efforts in AI alignment, a substantial body of research has focused on tuning methods to ensure that LLM outputs are helpful, honest, and harmless to humans~\citep{askell2021general, bai2022training, ouyang2022training}.
However, there remains a limited amount of detailed knowledge regarding the extent to which affective or perceptual evaluations of images are aligned with human values~\citep{shen2024bidirectional}.
In particular, aesthetic evaluation lacks clearly defined correct answers, unlike domains such as mathematics, and varies widely across individuals; therefore, it remains unclear to what extent such evaluations can be learned from large-scale textual data as a form of collective intelligence. 

Several studies have recently begun to investigate the effectiveness and limitations of LLMs in the task of predicting aesthetic evaluations of images. 
For example, it has been shown that LLMs can achieve aesthetic prediction performance exceeding chance level without requiring task-specific training~\citep{abe2025harnessingthepowerofllms}.
Moreover, in a study where multiple LLMs were prompted to generate aesthetic evaluations conditioned on demographic attributes, it was found that stereotype bias exists in the outputs, and that its magnitude varies depending on model size~\citep{li2025aesbiasbench}.
However, there have been few studies that examine in detail the internal process underlying aesthetic evaluation in LLMs. 
Human aesthetic evaluation is not formed solely based on objective image features; rather, it is influenced by internal processes within individuals~\citep{reber2004processing}, likely including emotions and bodily sensations involved in processing the visual stimulus. 
A study by Washizu et al.~\citep{washizu2025bodily} investigated these relationships in human aesthetic experience. 
In the study, more than 500 participants were presented with visual stimuli, and their aesthetic ratings were collected alongside measures of emotional categories and subjective bodily sensations, enabling an analysis of the relationships among these factors. 
Building on their work, the present study presents LLMs with comparable questions as those given to human participants, to investigate which aspects differentiate humans and LLMs in the process of aesthetic evaluation of visual stimuli. 

\section{Results}\label{sec2}
In this study, we compared the responses of human participants and multiple LLMs. 
The human behavioral data were obtained from~\citep{washizu2025bodily}.
In that study, human participants were asked to provide beauty, valence, and arousal scores, emotion labels, and reports of bodily sensations for 347 images selected from the PARA dataset~\citep{yang2022personalized}.
We posed the same set of questions to LLMs and collected their behavioral data. 
Specifically, each LLM was presented with a single image and asked to provide ratings of beauty, valence, and arousal on a 9-point Likert scale. 
In addition, for 32 emotion labels, the models were instructed to respond either using a \textit{ranking method}, in which they selected the top three labels, or a \textit{rating method}, in which they assigned a score on a 9-point scale to each label. 
Furthermore, for seven body parts, the models were asked to report the degree of association as an integer value. 
As part of the prompt design, we prepared conditions that varied by language (Japanese or English), as well as by generation setting: a \textit{deterministic setting}, in which a single deterministic output is produced, and a \textit{stochastic setting}, in which stochastic responses are generated multiple times and averaged. 

As LLMs, we examined frontier models as of June 2025 (GPT-4o~\citep{openai2024hello}, 
Claude 3.7 Sonnet~\citep{anthropic2025claude37}, and Gemini 2.0 Flash~\citep{google2024next}), 
hereafter referred to as GPT, Claude, and Gemini, respectively.
These models can be applied to a wide range of tasks in a zero-shot or few-shot manner--namely, without requiring additional training--through natural language instructions. 
They are also capable of accepting not only text but also images as input.

\subsection{Analysis 1: Relationship with Emotions}

First, prior to the main analysis of the relationship between beauty and emotion, we investigated biases in the distribution of emotions reported in the process of aesthetic evaluation of images for both humans and LLMs. 
Specifically, for the human evaluators, we defined the \textit{zero responses} for each emotion label as the number of images for which none of the participants selected that label, and computed this for all emotion labels. 
Similarly, the same procedure was applied to the LLM evaluators. 

\begin{figure}[H]
  \centering
  \includegraphics[width=0.8\columnwidth, trim=0 0 0 0, clip]{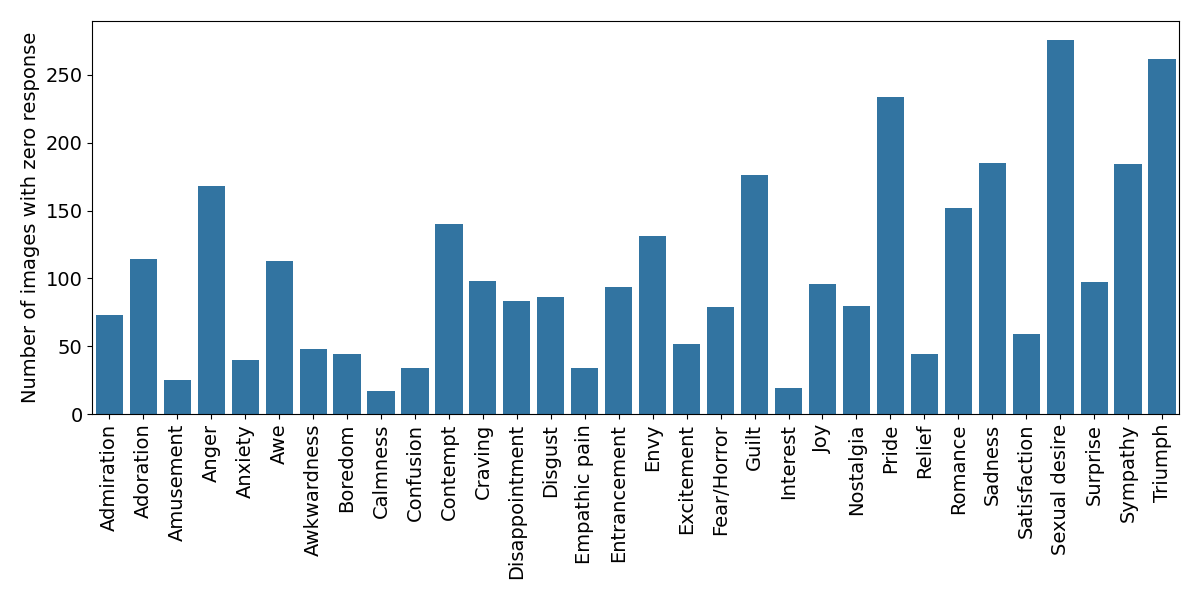}
  \caption[Number of images with zero response in human evaluations]{
  Number of images with zero responses in human evaluations, where the emotion never appeared among the top-3 responses.
  }
  \label{fig:zero_counts_human}
\end{figure}

Figure~\ref{fig:zero_counts_human} shows the distribution of zero responses in human evaluations. 
Emotional responses exhibit clear biases: for example, Amusement, Calmness, and Interest have fewer than 50 zero responses, indicating that for more than 300 images, at least one participant reported these emotions. 
In contrast, emotions such as Pride, Sexual desire, and Triumph have more than 200 zero responses, indicating that for many images, no participants reported these emotions. 

\begin{figure}[H]
  \centering
  \includegraphics[width=\columnwidth, trim=0 0 0 0, clip]{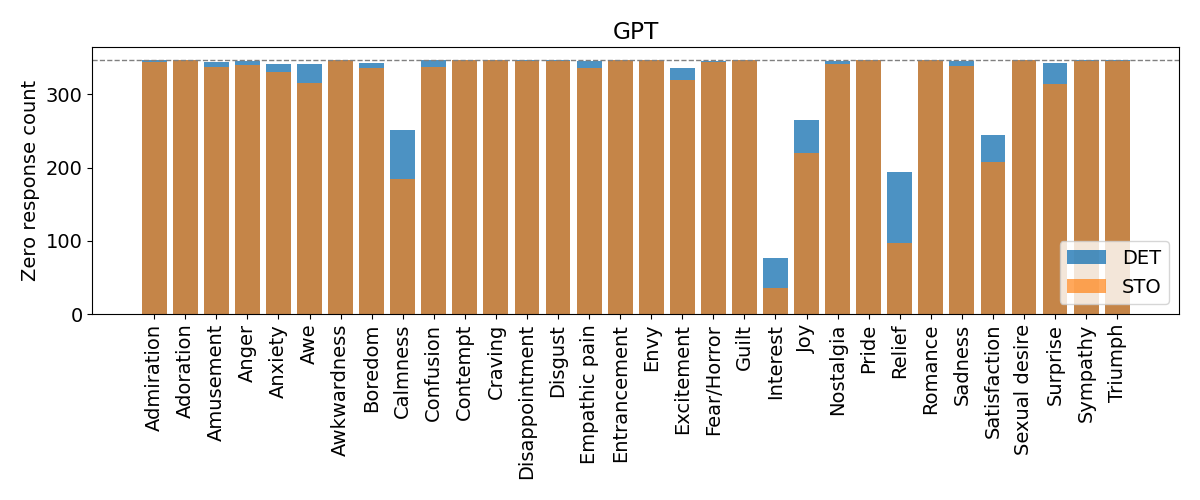}
  \includegraphics[width=\columnwidth, trim=0 0 0 0, clip]{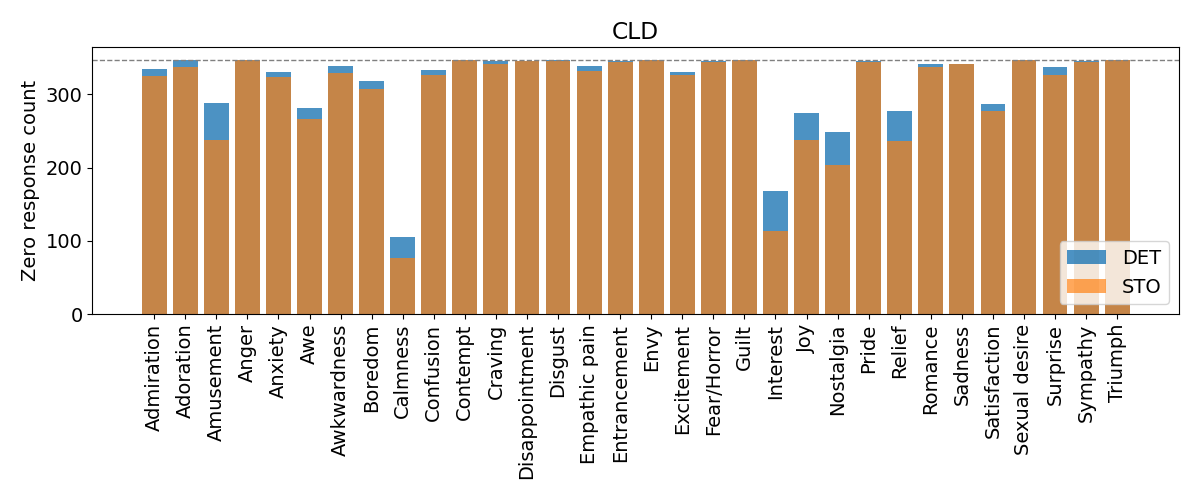}
  \includegraphics[width=\columnwidth, trim=0 0 0 0, clip]{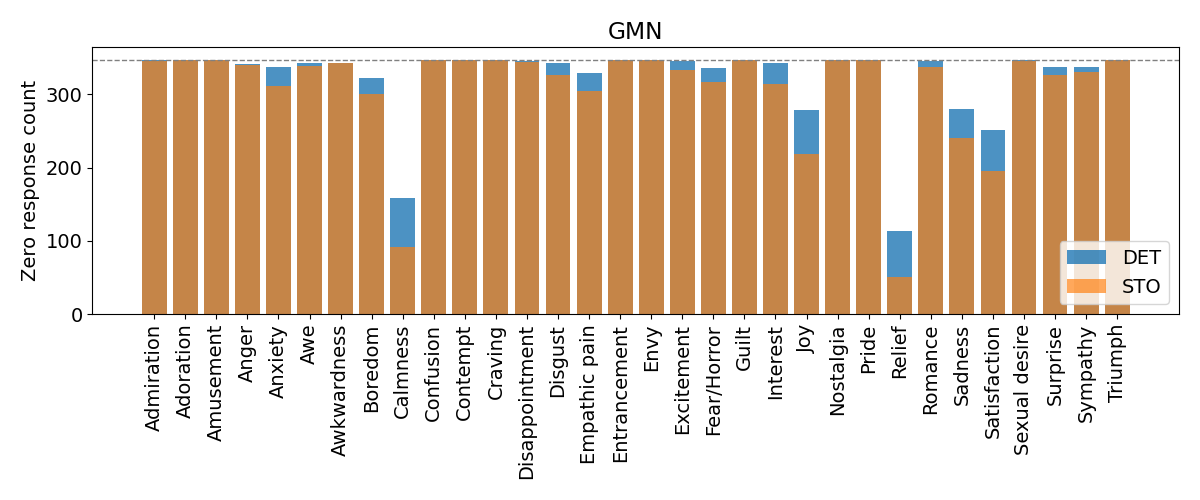}
  \caption[Number of images with zero response in AI evaluations]{
  Number of images with zero responses in AI evaluations, where the emotion never appeared among the top-3 responses.
  The abbreviations are as follows: DET/STO denote the determinism of the responses (Deterministic setting/Stochastic setting); GPT/CLD/GMN denote the language models (GPT/Claude/Gemini).
  The gray dashed line represents the total number of images (347).
  }
  \label{fig:zero_counts_ai}
\end{figure}

Figure~\ref{fig:zero_counts_ai} shows the distribution of zero responses in the evaluations under the ranking method by AI models. 
The distribution of emotional responses in AI is even more strongly biased than in humans. 
Under the deterministic setting, 
GPT more frequently produces responses in the order of Interest, Relief, and Satisfaction; 
Claude in the order of Calmness, Interest, and Nostalgia; 
and Gemini in the order of Relief, Calmness, and Satisfaction.
Under the stochastic setting, the frequency of certain responses increases substantially: 
for GPT, Relief, Calmness, and Joy; 
for Claude, Interest, Amusement, Nostalgia; 
and for Gemini, Calmness, Relief, and Joy.
On the other hand, there remain many emotion labels whose frequency of appearing among the top three hardly increases, despite the use of a stochastic setting with higher temperature and a larger number of calls.

\begin{figure}[H]
  \centering
  \includegraphics[width=\columnwidth, trim=0 0 0 0, clip]{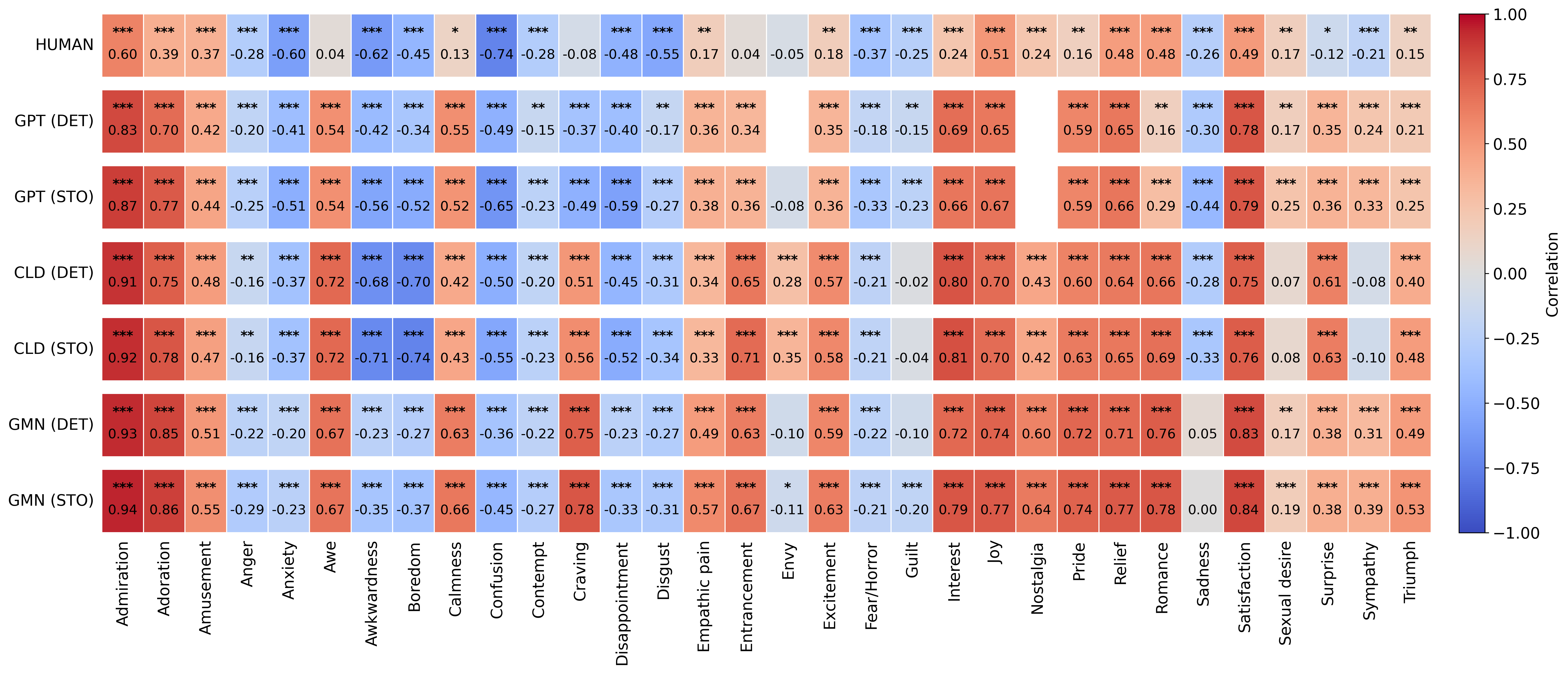}
  \caption[Spearman's rank correlation coefficients between the beauty and emotion intensity scores]{
    Spearman's rank correlation coefficients between the beauty and the emotion intensity scores across the seven evaluator conditions (HUMAN: using the average ratings from human evaluators; others: using scores from AI-based evaluators). 
    The abbreviations are as follows: DET/STO denote the determinism of the responses (Deterministic setting/Stochastic setting); GPT/CLD/GMN denote the language models (GPT/Claude/Gemini).
    To account for multiple comparisons, p-values were adjusted using the Benjamini–Hochberg correction procedure to control the false discovery rate within each evaluator condition. 
    Asterisks indicate statistical significance after the correction (*: $p < 0.05$, **: $p < 0.01$, ***: $p < 0.001$). 
    Blank entries indicate cases where the LLM responses collapsed, making the computation infeasible.
  }
  \label{fig:beauty_emotion_correlation}
\end{figure}

Next, we investigated the relationships between beauty, valence, and arousal scores and each emotion label. 
In the ranking method, where the top three emotion labels are selected, strong biases were observed in the labels produced by AI, resulting in many labels that were rarely selected. 
Therefore, in this analysis, we adopted the rating method, in which the intensity of each emotion label was evaluated.
For each evaluator, we computed correlations between beauty and emotion intensity scores across the 347 images. 
The results are shown in Figure~\ref{fig:beauty_emotion_correlation} \footnote{For the results of the correlation analysis between valence/arousal and emotion intensity scores, see the Supplementary Information section.}.

Overall, humans and AI are largely consistent in terms of the types of emotions associated with aesthetic evaluation. 
For example, positive emotions such as Admiration, Joy, and Satisfaction, which show strong positive correlations in humans, also exhibit strong positive correlations in AI. 
Conversely, ambiguous or negative emotions such as Confusion, Awkwardness, Anxiety, Disgust, and Disappointment, which show strong negative correlations in humans, likewise exhibit strong negative correlations in AI. 

On the other hand, there are also emotion categories for which discrepancies between humans and AI can be observed. 
For Awe and Entrancement, no significant correlation with beauty is observed in humans, whereas strong positive correlations are found across all six AI evaluator conditions. 
For Craving, no significant correlation with beauty is observed in humans; however, GPT shows a strong negative correlation, while Claude and Gemini show strong positive correlations. 
For Surprise and Sympathy, humans exhibit significant negative correlations with beauty, whereas in the AI evaluator conditions, either no correlation or strong positive correlations are observed. 

\subsection{Analysis 2: Relationship with Bodily Sensations}
\begin{figure*}[htbp]
  \centering
  \begin{subfigure}[t]{0.34\textwidth}
    \includegraphics[width=\linewidth, trim=0 0 75 0, clip]{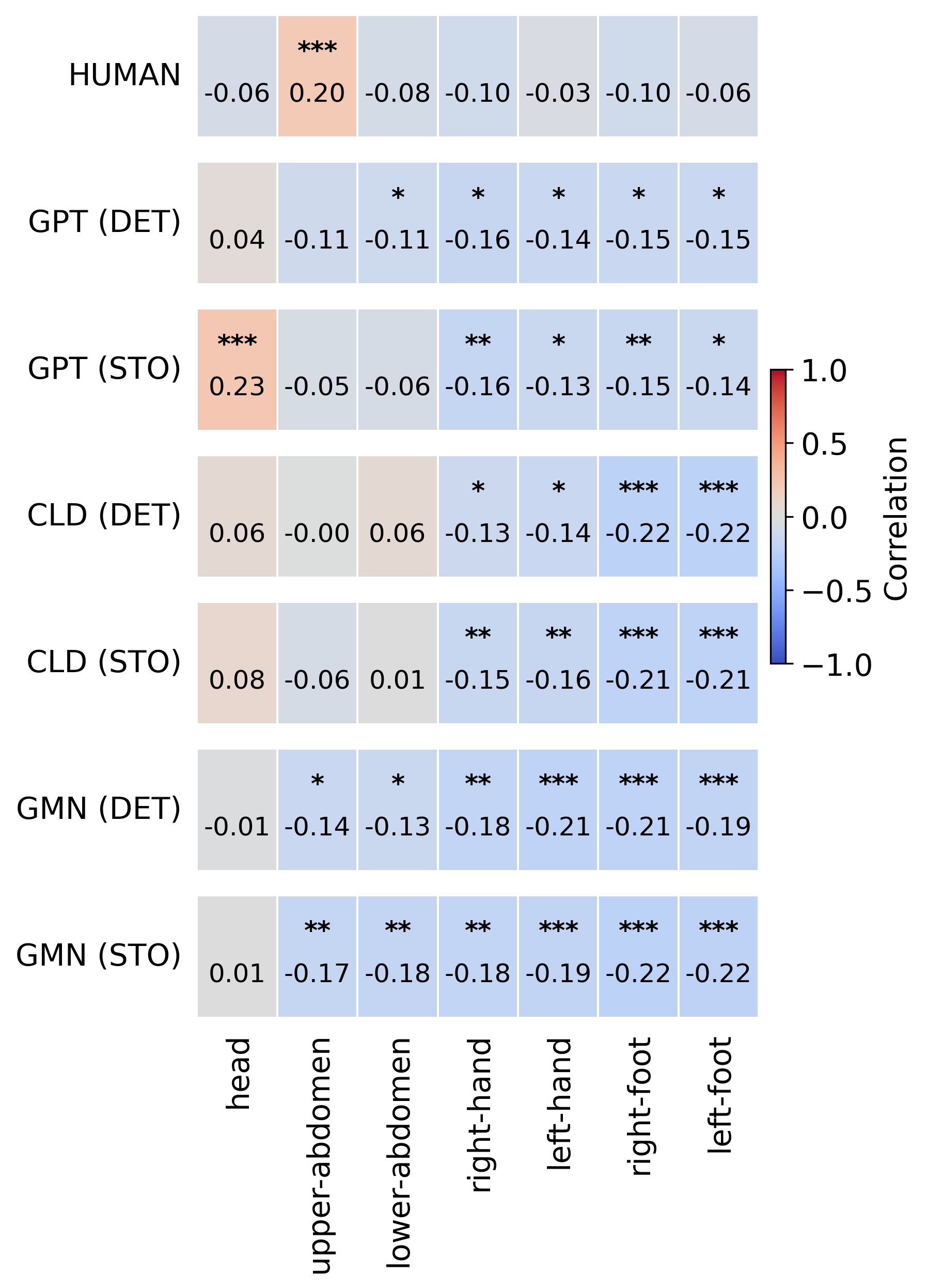}
    \caption{Beauty}
    \label{fig:beauty_bodymap_correlation}
  \end{subfigure}
  \hfill
  \begin{subfigure}[t]{0.2585\textwidth}
    \includegraphics[width=\linewidth, trim=85 0 75 0, clip]{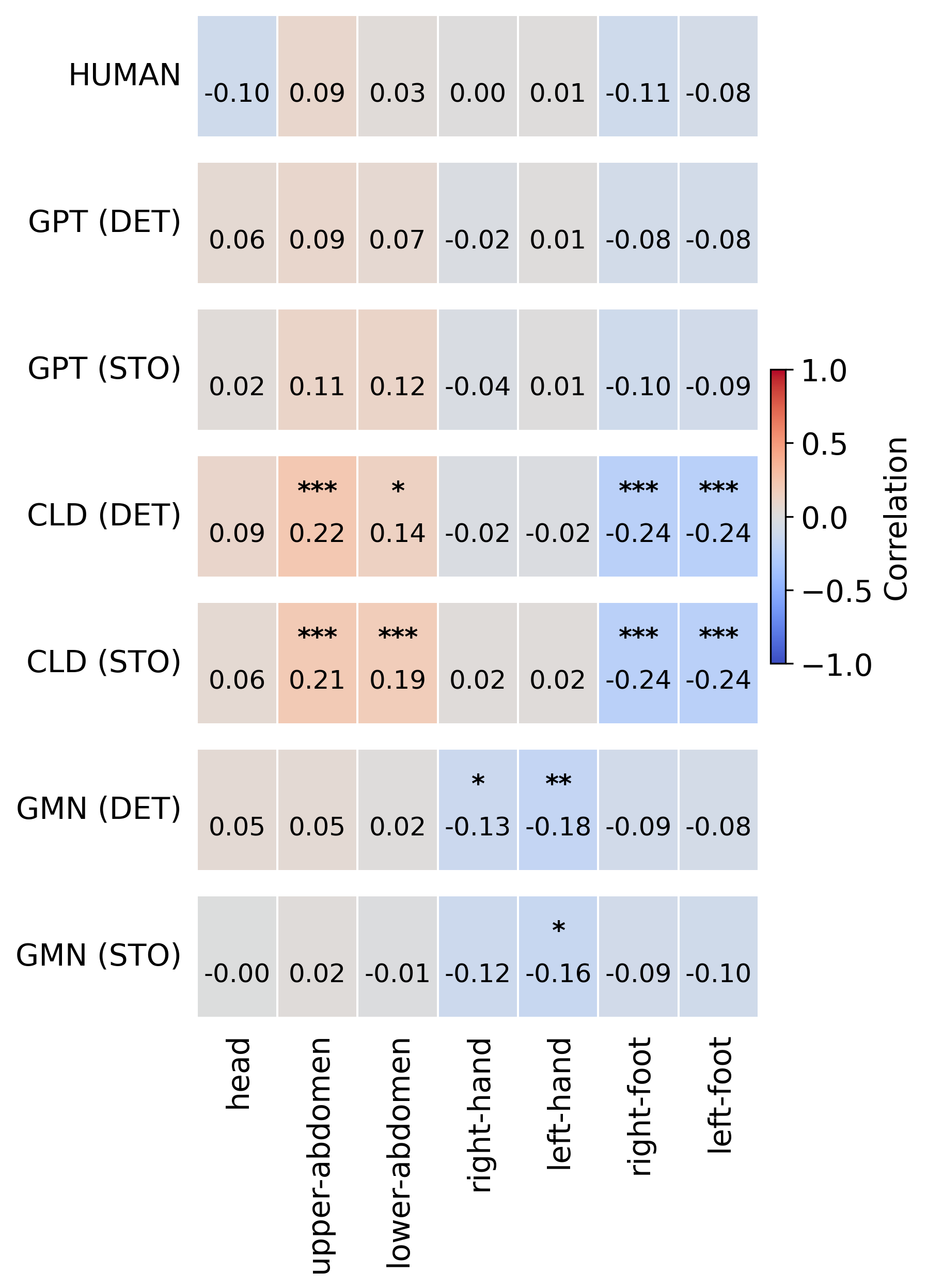}
    \caption{Valence}
    \label{fig:valence_bodymap_correlation}
  \end{subfigure}
  \hfill
  \begin{subfigure}[t]{0.3305\textwidth}
    \includegraphics[width=\linewidth, trim=85 0 0 0, clip]{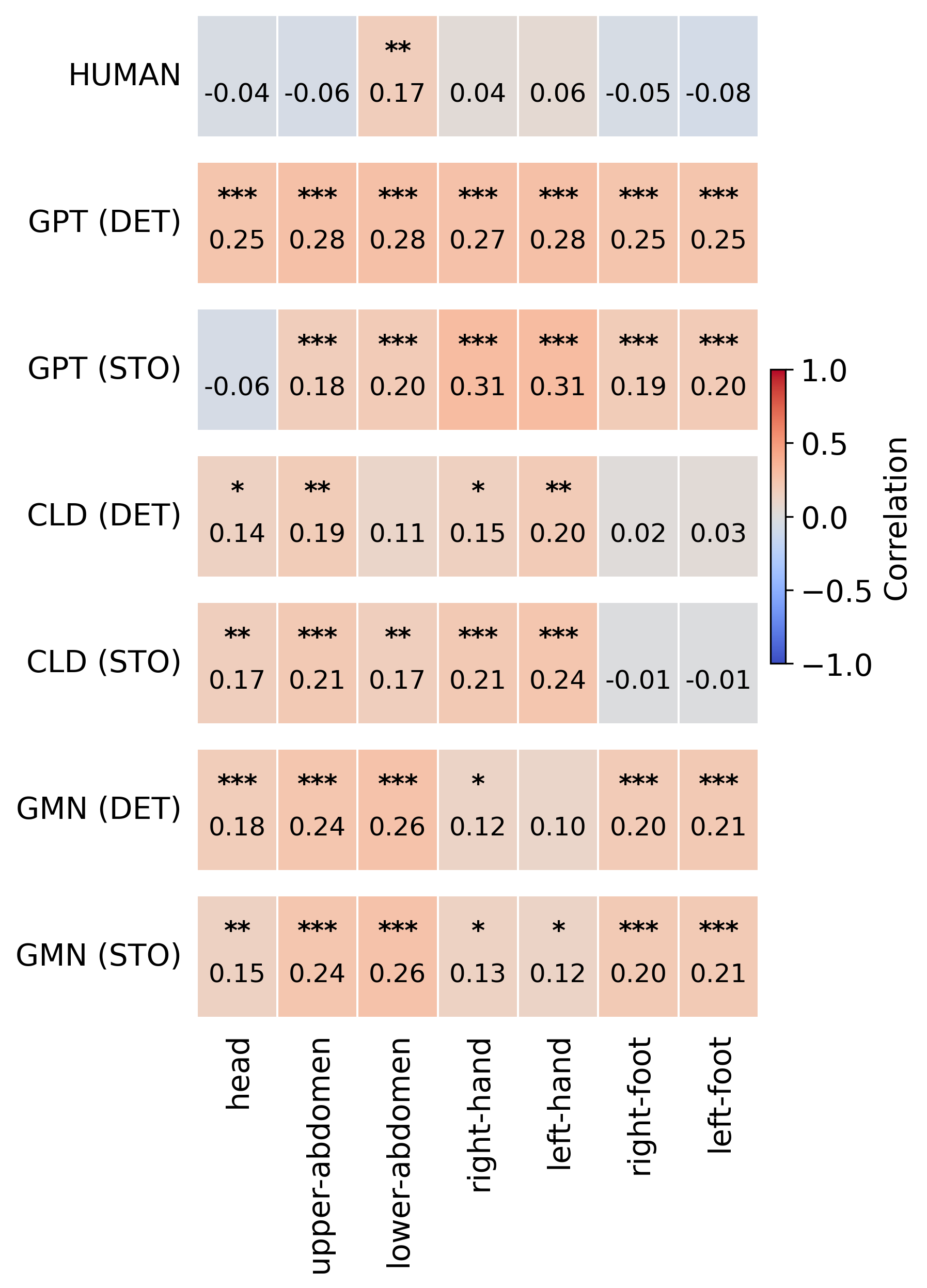}
    \caption{Arousal}
    \label{fig:arousal_bodymap_correlation}
  \end{subfigure}
  \caption[Spearman's rank correlation coefficients between the beauty/valence/arousal and bodily sensation scores]{
  Spearman's rank correlation coefficients between the beauty/valence/arousal scores and the bodily sensation scores across the seven evaluator conditions (HUMAN: using the average ratings from human evaluators; others: using scores from AI-based evaluators). 
  The abbreviations are as follows: DET/STO denote the determinism of the responses (Deterministic setting/Stochastic setting); GPT/CLD/GMN denote the language models (GPT/Claude/Gemini).
  To account for multiple comparisons, p-values were adjusted using the Benjamini-Hochberg correction procedure to control the false discovery rate within each evaluator condition. 
  Asterisks indicate statistical significance after the correction (*: $p < 0.05$, **: $p < 0.01$, ***: $p < 0.001$). 
  }
  \label{fig:bodymap_correlations}
\end{figure*}

We investigated the relationships between beauty, valence, and arousal and bodily sensations. 
For each evaluator, we computed correlations across the 347 images. The results are shown in Figure~\ref{fig:bodymap_correlations}.

Unlike the relationship between beauty and emotions, the relationship between beauty and bodily sensations differs substantially between humans and AI. 
In humans, beauty shows a strong positive correlation with sensations in the upper abdomen, while no significant correlations are observed for other body parts. 
In contrast, for many AI evaluators, no positive correlation is observed with upper abdominal sensations; instead, strong negative correlations are observed with sensations in the hands and feet. 
Furthermore, in GPT (stochastic setting), a strong positive correlation is observed with sensations in the head. 

Similarly, we investigated the relationships between valence and arousal, and bodily sensations. 
In humans, no significant correlations are observed between valence and any body parts, whereas in AI evaluators, the results vary depending on the model. 
For arousal, the differences between humans and AI are more pronounced. 
In humans, a strong positive correlation is observed only with sensations in the lower abdomen, whereas in AI, many conditions exhibit strong positive correlations across the entire body. 

Next, we examine the overall tendencies of bodily sensations for each AI model. 
In GPT, arousal shows positive correlations with many body parts, and beauty shows negative correlations with the hands and feet. 
In Claude, arousal shows positive correlations with body parts except the feet, and valence shows positive correlations with the upper and lower abdomen; in contrast, beauty shows negative correlations with the hands and feet, and valence shows a negative correlation with the feet. 
In Gemini, arousal shows positive correlations with many body parts, while beauty shows negative correlations with all body parts except the head. 
Thus, these results indicate that the relationships between body parts and beauty, valence, and arousal differ across AI models; however, all of them show substantial discrepancies from human patterns.

\subsection{Analysis 3: Aesthetic Evaluation Scores}

\begin{figure}[H]
  \centering
  \includegraphics[width=\columnwidth, trim=0 0 0 0, clip]{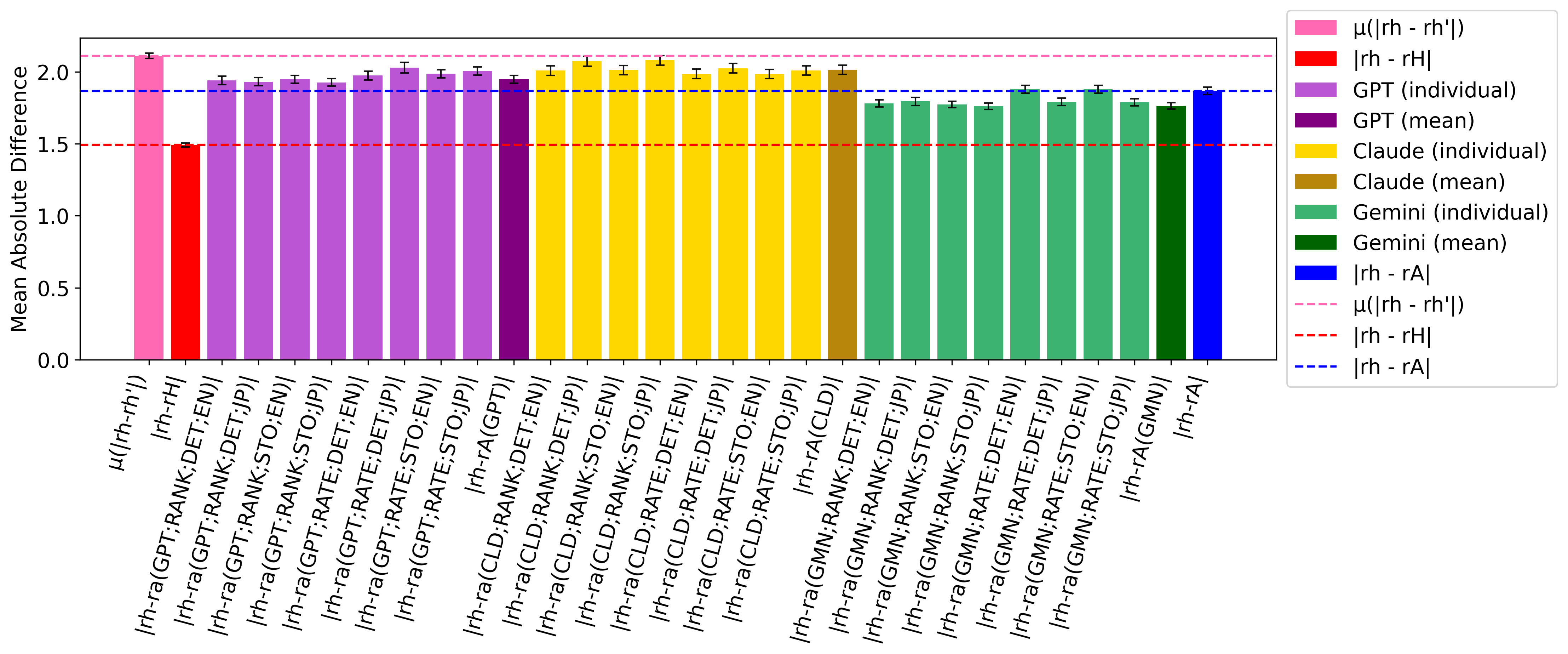}
  \caption[Differences from individual human evaluator]{
    Differences from individual human evaluators ($r_h$). 
    Values are first averaged across multiple raters of the same image, and then averaged across all 347 images. 
    Error bars indicate the standard error of the mean.
    The pink bar represents the variability among human evaluators ($\mu(|r_h - r_{h'}|)$); the red bar represents the deviation of an individual human evaluator from the human average ($|r_h - r_H|$); and the blue bar represents the deviation of an individual human evaluator from the overall average of all AI-based evaluators ($|r_h - r_A|$). 
    The other bars in purple, yellow, and green represent the deviations between an individual human evaluator and GPT/Claude/Gemini ($|r_h - r_a|$ or $|r_h - r_A|$; the darker colors indicate the averages of each LLM's evaluations). 
    The abbreviations used in the AI evaluations are as follows: RANK/RATE denote the emotion evaluation method (Ranking method / Rating method); DET/STO denote the determinism of the responses (Deterministic setting/Stochastic setting); GPT/CLD/GMN denote the language models (GPT/Claude/Gemini); and EN/JP denote the prompt language (English/Japanese).
  }
  \label{fig:abs_diff_plot_rh}
\end{figure}

We conducted an analysis to clarify the differences between humans and AI in the aesthetic evaluation (beauty) scores themselves.
Figure~\ref{fig:abs_diff_plot_rh} presents a plot comparing divergence between the ratings of individual human evaluators, denoted as $r_h$, and those of various other evaluators.
In this graph, a value of zero on the vertical axis indicates complete agreement with an individual's rating.
The condition closest to this situation is the red bar, $|r_h - r_H|$, indicating that the mean rating of human evaluators is the closest. 
In contrast, the pink bar, $\mu(|r_h - r_{h'}|)$, shows the largest values on the vertical axis, indicating that the divergence between individuals is greater than the divergence between an individual and any AI evaluator (i.e. $|r_h - r_a|$ and $|r_h - r_A|$).
Among the AI models, differences due to response format, determinism settings, and language are relatively small. 
When comparing across models, Gemini is found to be the closest to individual human ratings.

\begin{figure}[H]
  \centering
  \includegraphics[width=\columnwidth, trim=0 0 0 0, clip]{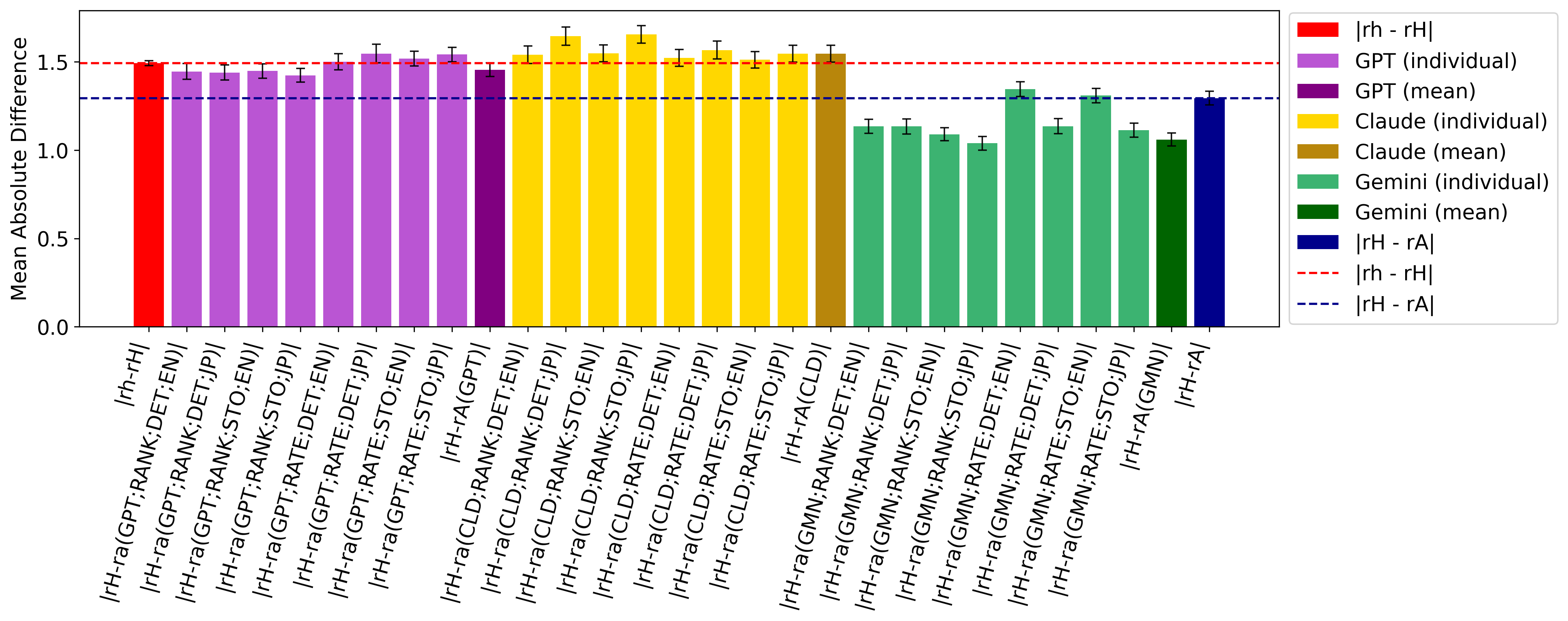}
  \caption[Differences from the human average]{
  Differences from the human average ($r_H$). 
  Values are first averaged across multiple raters of the same image, and then averaged across all 347 images. 
  Error bars indicate the standard error of the mean.
  The red bars represent the deviation of an individual human evaluator from the human average ($|r_h - r_H|$);
  the blue bars represent the deviation of the human average from the overall average of all AI-based evaluators ($|r_H - r_A|$).
  The other bars in purple, yellow, and green represent the deviations between the human average and GPT/Claude/Gemini ($|r_H - r_a|$ or $|r_H - r_A|$; the darker colors indicate the averages of each LLM's evaluations). 
  The abbreviations used in the AI evaluations are as follows: RANK/RATE denote the emotion evaluation method (Ranking method/Rating method); DET/STO denote the determinism of the responses (Deterministic setting/Stochastic setting); GPT/CLD/GMN denote the language models (GPT/Claude/Gemini); and EN/JP denote the prompt language (English/Japanese).
  }
  \label{fig:abs_diff_plot_rLargeH}
\end{figure}

Figure~\ref{fig:abs_diff_plot_rLargeH} shows a plot comparing the degree of divergence between the mean human rating, $r_H$, and the ratings of various other evaluators. 
Among the LLMs, Gemini is the closest to the human mean. 
It is also observed that $|r_H - r_A|$ is smaller than $|r_h - r_H|$.

\begin{figure}[H]
  \centering
  \includegraphics[width=\columnwidth, trim=0 0 0 55, clip]{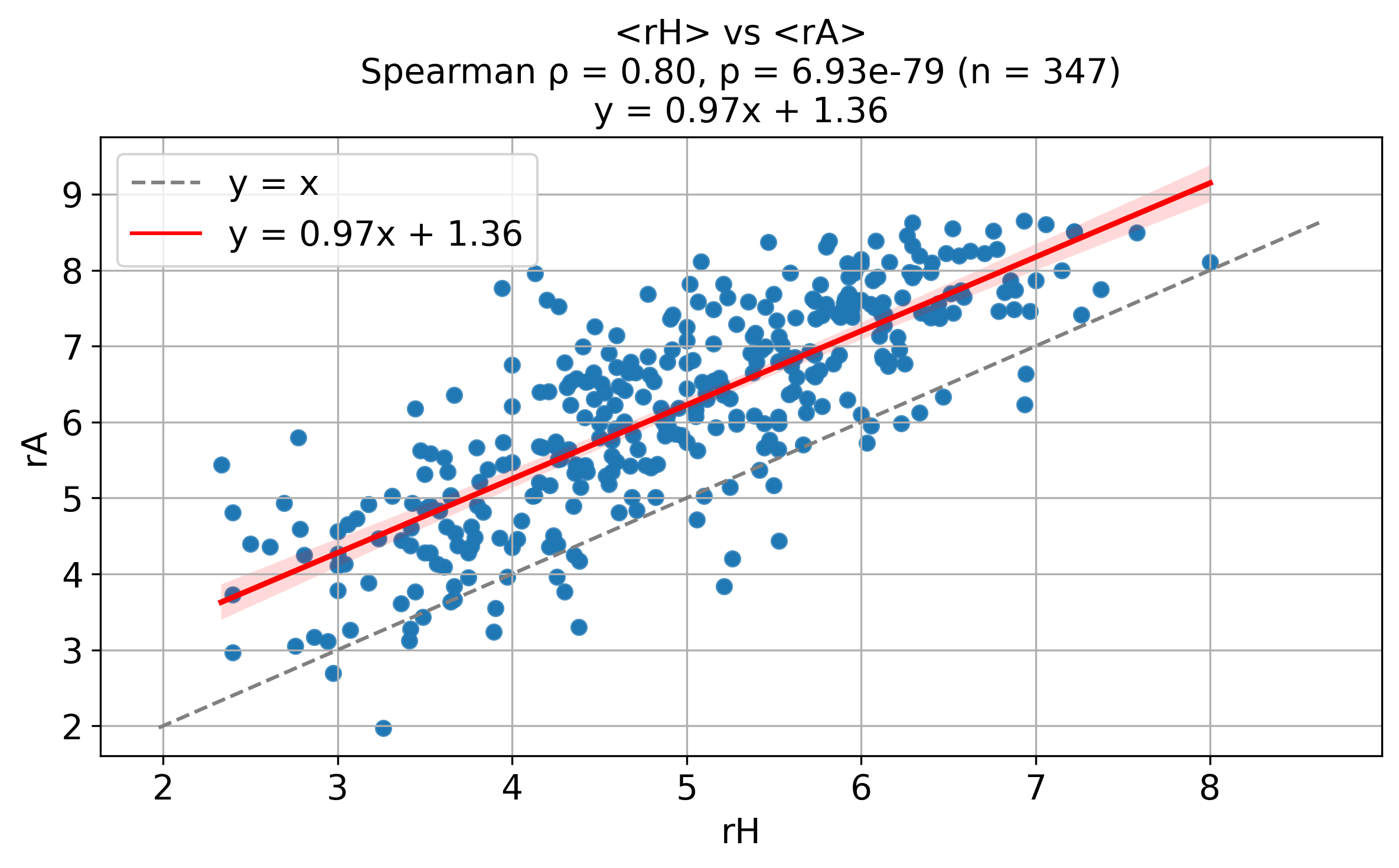}
  \caption[Relationship between the human average and the AI average]{
  Relationship between the human average ($r_H$) and the AI average ($r_A$), with $r_H$ on the x-axis and $r_A$ on the y-axis. 
  A total of 347 images are included, with the regression line and the 95\% confidence interval shown as a red line and red shaded area, respectively.
  }
  \label{fig:scatter_mean__Beauty_vs_mean_aesthetics_all__Q3-beauty}
\end{figure}

Figure~\ref{fig:scatter_mean__Beauty_vs_mean_aesthetics_all__Q3-beauty} plots the relationship between the mean human rating, $r_H$, and the mean AI rating, $r_A$, across all 347 images, with $r_H$ on the x-axis and $r_A$ on the y-axis. 
The fitted regression line, $y=0.97x+1.36$, is nearly parallel to the identity line $y=x$.
Mean AI ratings tend to be higher than mean human ratings by a roughly constant offset (approximately 1.3 to 1.4 points).

\subsection{Analysis 4: Image Features}

\begin{figure}[H]
  \centering
  \includegraphics[width=\columnwidth, trim=0 0 0 0, clip]{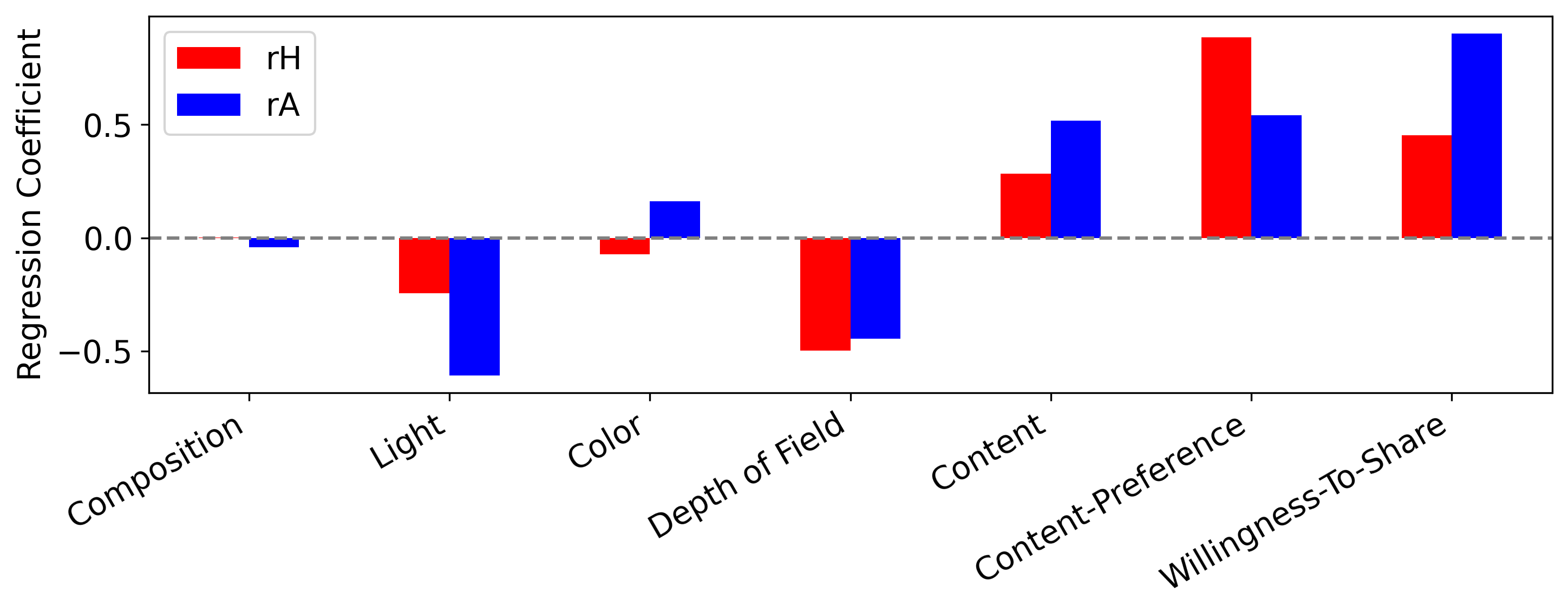}
  \caption[Regression coefficients of image features in linear regression]{
    Regression coefficients of image features in linear regression with the human average ($r_H$) and the AI average ($r_A$) as dependent variables.
  }
  \label{fig:coef_comparison}
\end{figure}

We investigated how humans and AI differ in the image features they prioritize. 
Figure~\ref{fig:coef_comparison} shows the coefficients corresponding to each image feature in two linear regression models, where the mean human rating $r_H$ and the mean AI rating $r_A$ are used as target variables. 
The magnitudes of the coefficients are compared after standardizing each image feature. 
For humans, Content-Preference has the largest contribution, whereas for AI, Willingness-To-Share has the largest contribution. 
Both of these features are high-level image features involving subjective judgments. 

\begin{figure}[H]
  \centering
  \includegraphics[width=\columnwidth, trim=0 0 0 0, clip]{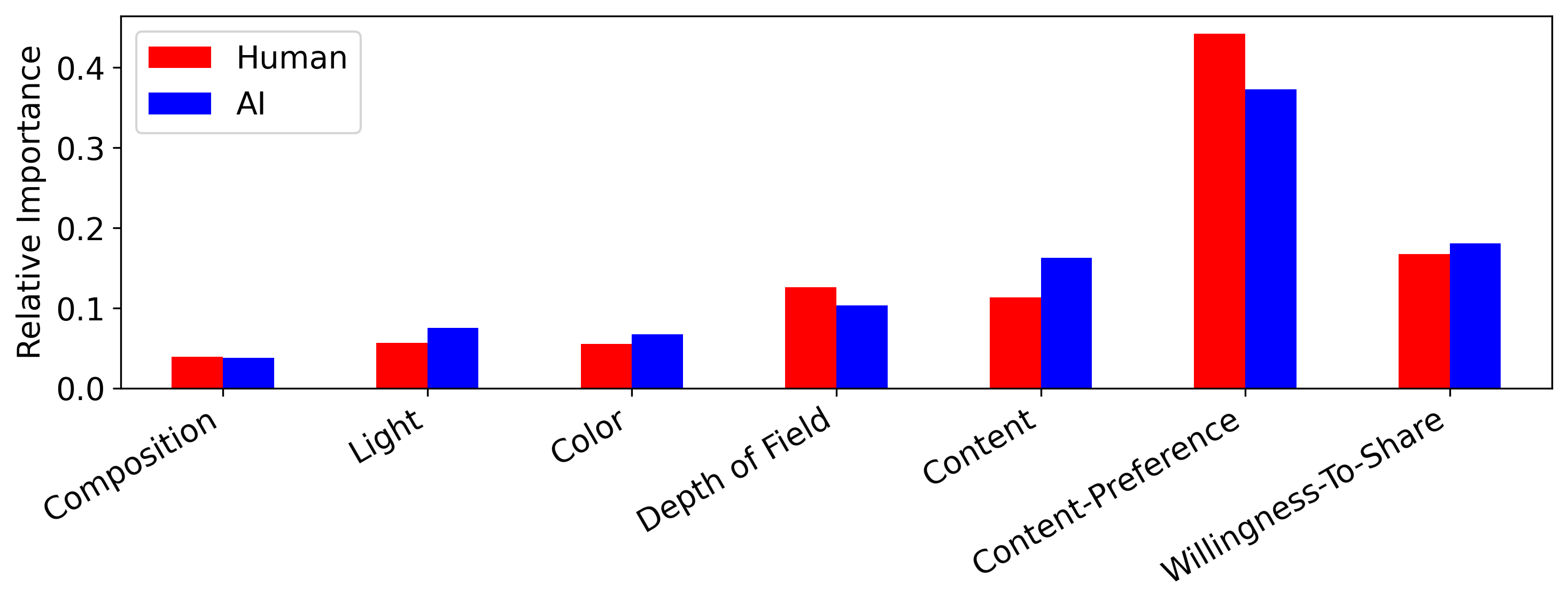}
  \caption[Relative importance of image features in SHAP]{
    Relative importance of image features in SHAP. 
    Using two random forest regression models with the human average ($r_H$) and the AI average ($r_A$) as dependent variables, SHAP values were computed for each image. 
    The absolute values of the SHAP scores were averaged across images, and then normalized across features to obtain their relative importance.
  }
  \label{fig:relative_importance}
\end{figure}

To further examine feature contributions in the presence of potential multicollinearity among image features, we additionally conducted an analysis using SHapley Additive exPlanations (SHAP)~\citep{lundberg2017shap}.
Figure~\ref{fig:relative_importance} compares the magnitudes of feature importance obtained from random forest regression models trained to predict $r_H$ and $r_A$, using the mean absolute SHAP values across all 347 images for each feature.
The coefficients of determination for the random forest regression models were $0.922$ for the $r_H$ model and $0.932$ for the $r_A$ model. 
Since both values are close to $1$, this indicates that the models achieve high predictive accuracy. 
For both models, Content-Preference is identified as the most important feature, followed by Willingness-To-Share. 
Consistent with the findings from the linear regression analysis, it is again confirmed that both humans and AI place greater emphasis on subjective, high-level features in aesthetic evaluation. 

\section{Discussion}\label{sec3}

\subsection{Analysis 1: Relationship with Emotions}
We examined the relationship between aesthetic evaluation and emotions in humans and AI. 
First, AI exhibits a stronger bias in the distribution of emotion labels across the 347 images than humans when using the ranking method.
However, it should be noted that in the human evaluator condition, responses from 500 participants were used, whereas in the AI evaluator condition, even under the stochastic setting (temperature = 0.7), responses were obtained from only five calls. 
Thus, the difference may partly reflect the larger sample size in the human condition. 
Additionally, even under the stochastic setting, the responses reflect only repeated samples from the same model, which may also contribute to the observed difference. 
However, even within the same model, shifting the generation setting from deterministic to stochastic increases the frequency of already common emotion labels, such as Interest, Calmness, Relief, while responses for less frequent labels, such as Guilt, Sexual desire, and Triumph, show little increase.

In light of these observations, the main part of Analysis 1 focused on the rating method in order to compute correlations between beauty and emotion intensity scores across all images. 
According to the results of the Spearman correlation analysis, the relationship between beauty scores and emotion categories shows similar tendencies between humans and AI for major emotions such as Joy, Satisfaction, Disgust, and Disappointment. 
However, notable discrepancies between humans and AI are observed for emotions such as Awe, Entrancement, Craving, Surprise, and Sympathy. 

For Awe, humans tend to associate it with negative valence, like fear or anxiety, whereas AI may process it as being directly associated with beauty, as often discussed in the aesthetics literature~\citep{pelowski2017move}.
For Craving, humans tend to distinguish bodily or physiological desires from aesthetic appreciation, whereas AI may fail to make such a distinction. 
A similar tendency is observed for Entrancement. 
For Surprise, although it is not necessarily perceived as positive by humans, AI may associate it positively with beauty as an indicator of novelty (a key aspect in the evaluation of creativity~\citep{boden2004creative}).
For Sympathy, humans tend to associate it with negative valence, similar to sadness, whereas AI may reflect the moral beauty of sympathy, or alternatively, may not experience strong sympathy in the same way as humans, resulting in a reduced negative influence.

\subsection{Analysis 2: Relationship with Bodily Sensations}
We examined the relationship between aesthetic evaluation and bodily sensations in humans and AI. 
Spearman correlation analysis shows that the relationship between beauty and bodily sensations differs substantially between humans and AI, unlike the relationship between beauty and emotions. 
In humans, only sensations in the upper abdomen show a strong positive correlation with beauty, whereas in AI, no positive correlation is observed for the upper abdomen; instead, negative correlations are primarily observed with the hands and feet. 
In addition, the relationship between arousal and bodily sensations further reveals a substantial discrepancy between humans and AI. 
In humans, only sensations in the lower abdomen show a strong positive correlation with arousal, whereas in AI, correlations are observed with multiple parts across the entire body. 
These results suggest differences in the internal processes underlying aesthetic evaluation of images between humans and AI. 
In humans, beauty and arousal appear to be associated with interoceptive sensations--particularly the intensity of sensations in the upper or lower abdomen--whereas in AI, beauty appears to be associated with a relative attenuation of sensations in the hands and feet, and arousal with increased sensations across the body. 

In AI, specifically in LLMs, these differences may be attributed to the lack of human-like embodiment (e.g., the absence of a human bodily structure including internal organs), 
that is, the fact that they do not acquire knowledge as physically existing entities in the real world. 
Additionally, slight differences in the method of reporting bodily sensations between humans and AI, due to experimental constraints, may also have influenced the results.
Human participants reported bodily sensations by clicking on locations on a body map image corresponding to their perceived sensations, whereas AI evaluators were provided with seven predefined body parts and asked to report the intensity of association for each as an integer value.

\subsection{Analysis 3: Aesthetic Evaluation Scores}
First, focusing on aesthetic evaluation (beauty) scores, we examined the differences between human evaluators and AI evaluators. 
In terms of deviation from individual human evaluators, the smallest deviation was observed for the human mean, followed by the AI mean, while the largest deviation was observed between individuals. 
This indicates that, from the perspective of a given human evaluator, AI is closer than other individual human evaluators, but still farther than the human average. 
Furthermore, in terms of deviation from the mean of human evaluators, the deviation of the AI mean was smaller than that of individual human evaluators. 
This result suggests that AI more closely approximates the population average of human evaluations rather than those of any specific individual, and also emphasizes the existence of individual differences in human evaluations. 
In addition, a linear regression of mean AI ratings on mean human ratings indicates a tendency for AI ratings to be higher than human ratings by a roughly constant offset (about 1.3--1.4 points on a 1--9 Likert scale). 

These results suggest that AI can approximate the average aesthetic evaluations of a human population to a certain extent. 
On the other hand, in terms of divergence from individual aesthetic evaluations, AI does not outperform the human average. 
In the present experimental setup, AI is not set up to produce different evaluations specific to individual human evaluators; therefore, it is inherently difficult for AI outputs to be closer to each individual rating than the human mean, unless the distribution of human ratings for a given image is multimodal and the AI happens to align with one of the modes. 
Therefore, these results do not rule out the possibility that AI could learn and predict individual aesthetic evaluations. 
Rather, they provide a benchmark for future work on modeling and predicting individual differences in aesthetic evaluation.

\subsection{Analysis 4: Image Features}
Finally, we investigated how humans and AI differ in the image features they prioritize during aesthetic evaluation.
We conducted two analyses: a comparison of regression coefficients in linear regression and a comparison of feature importance using SHAP in random forest regression. 
The results indicate that, for both humans and AI, subjective and high-level image features--such as preference for the content and the willingness to share the image with others--have a stronger influence on aesthetic evaluation than low-level features such as composition, light, and color. 

From the results of Analysis 4, humans and AI appear to be largely similar in terms of the image features they prioritize. 
However, in this analysis, the target variable was not the rating of individual human evaluators, but rather their average. 
When considered at the level of individual human evaluators, it is expected that attention to image features varies across individuals.
Indeed, in supplementary analyses examining the interaction between personality traits and image features, it was found that the influence of specific image features on beauty ratings changes depending on personality traits. 
Accordingly, modeling individual aesthetic evaluations may require AI systems to adapt their sensitivity to image features depending on the individual.

\subsection{Implication 1: Interoceptive World Model}
Taniguchi et al.~\citep{taniguchi2025generative} note that there have been multiple reports suggesting that today's LLMs possess knowledge that models the structure of the world--so-called world models--and propose a classification into two types. 
The first is a Type 1 world model, defined as "A subjective, internal model that an agent learns through its own sensorimotor interactions with the environment to predict future states and plan actions."
The second is a Type 2 world model, defined as "An objective, structured representation of knowledge about the world, its entities, and their relations, which may not be tied to a single agent's direct experience."
Based on this distinction, they frame a paradox at the center of ongoing debates on world models in LLMs: although current LLMs, lacking physical embodiment, would be expected to have difficulty acquiring Type 1 world models, they nevertheless appear to possess Type 2 world models. 
In response, they propose the hypothesis that human language emerges from interactions within society, in which individuals as embodied agents attempt to construct meaning, and that it functions as an externalized representation of an aggregated world model. 
By learning from such language, LLMs inherit the structure of this world model, which in turn makes them appear to possess knowledge about the world. 
They further argue that LLMs do not reconstruct the internal world models of individual humans, but rather predict a collective, aggregated abstraction of these representations. 

In this study, we aim to provide an extension to this hypothesis. 
According to Taniguchi et al., while current LLMs without embodiment are not capable of predicting the internal representations of individual humans, they should nevertheless be able to reproduce the collective structure of such internal representations given sufficiently large-scale data from human populations. 
However, in our analysis of the relationships between beauty or arousal and bodily sensations (Analysis 2), substantial discrepancies were observed when comparing LLM behavior with the averages across multiple human participants. 
A similar phenomenon has been reported in prior work by Xu et al.~\citep{xu2025largelanguagemodelssensorimotor}, which compared subjective evaluations of word concepts between humans and LLMs, showing that similarity to humans is lower for sensory and motor concepts than for non-sensorimotor concepts. 
While Taniguchi et al. emphasize sensorimotor interactions with the environment, we argue that it is important to explicitly distinguish between aspects of sensation that humans can precisely express in language and those that cannot. 
Taniguchi et al. argue that even LLMs, which lack a Type 1 world model (defined as a subjective and internal model), can acquire a Type 2 world model (i.e., an aggregated structure) through language. 
However, we suggest that certain aspects, particularly interoceptive bodily sensations, remain difficult to acquire through language alone, even when such language emerges from interactions among many embodied agents. 
We conceptualize these aspects as an \textit{Interoceptive World Model}. 
Here, the notion of world encompasses not only the external environment but also the internal state of the body. 
Furthermore, the interoceptive world model does not fall within either Type 1 or Type 2, but rather represents a distinct categorization of world models. 

Exteroceptive senses play an important role in interactions with the external environment and in somatic motor activity. 
Inputs to these sensory systems are derived from outside the body and are based on objective information that is, in general, observable by others. 
In contrast, interoceptive senses reflect the physiological condition of the internal body and contribute to the maintenance of homeostasis. 
These sensory inputs originate within the body (e.g., internal organs and the autonomic nervous system), do not involve direct physical interaction with the external environment, and typically constitute subjective information that is not observable by others~\citep{craig2003interoception, connell2018interoception}.
Therefore, compared to exteroceptive senses, interoceptive senses may exhibit a lower degree of consistency across individuals in how linguistic expressions map onto underlying phenomena, and methods for investigating them may still be underdeveloped. 
Supporting this view, prior studies suggest that the ability to accurately perceive interoceptive signals is associated with the richness of emotional expression~\citep{suzuki2023influence}.
Furthermore, alexithymia has been reported to be positively correlated with confusion in interoceptive sensations and negatively correlated with interoceptive accuracy~\citep{bael2024systematic}.
These findings suggest that there are substantial individual differences in the processing of perceiving interoceptive signals and translating them into linguistic emotional expressions. 
Moreover, it has been shown that objective interoceptive accuracy and subjectively reported interoceptive sensitivity do not necessarily align~\citep{garfinkel2015knowing}, suggesting that accurately perceiving and verbalizing one's own physiological state can be difficult for some individuals. 
Furthermore, methods for measuring interoception have been developed relatively recently, and it is possible that sufficiently high-quality data capturing shared structures across individuals have not yet been adequately accumulated~\citep{desmedt2023newmeasures}.

In summary, due to limitations in perceptual accuracy, the ability to verbalize internal states, and the degree of inter-individual consistency in meaning, knowledge related to interoceptive sensations is unlikely to be fully captured as a Type 2 world model, even with large-scale linguistic datasets. 
This distinction between exteroceptive and interoceptive processes suggests that, in order to build AI that is truly aligned with human perception, it is necessary to incorporate not only large-scale language and image data but also interoceptive information--often overlooked--as part of the training data. 
According to Connell et al.~\citep{connell2018interoception}, interoceptive processes play an important role in the perceptual grounding of both concrete and especially abstract concepts. 
Therefore, AI systems that lack such interoceptive capacities may exhibit forms of grounding that differ from, or are inferior to, those of humans.

Furthermore, in the context of beauty, recent studies (e.g.,~\citep{tschacher2012physiological, cabbai2024emotion}) have shown that physiological responses such as heart rate and skin conductance, as well as their subjective sensations, are closely related to aesthetic experience. 
However, current AI systems do not model such complex physiological processes or interoceptive signals, which may contribute to the observed discrepancies between humans and AI. 
Conversely, it is also possible that noise arising from internal bodily processes--and the resulting inherent unpredictability--constitutes a distinctive feature of human aesthetic evaluation. 
To what extent such aspects can be modeled in AI remains an open question.

A natural question, then, is how AI could acquire an interoceptive world model comparable to that of humans. 
One possible approach is to collect large-scale physiological and bodily sensation data by equipping the human body with extensive sensing devices and using such data for training AI systems. 
It may also be possible to construct detailed simulation models of the human body based on the data and allow AI to observe a wide range of phenomena within simulated environments, thereby facilitating the development of human-like world models (e.g.,~\citep{kim2022simulating}). 
Another possible approach is to employ physical robotic systems that mimic the internal structure of the human body and enable them to observe and learn from diverse phenomena in the real world. 
However, these approaches face several limitations, including the significant physical burden imposed on human participants, the existence of aspects of the real world that are difficult to capture (such as noise, continuous-valued dynamics, and irreversible processes like material degradation), and the practical challenges of implementation. 
Ultimately, if the goal is to enable AI to perceive subjective experiences comparable to those of humans, one possible direction may be a human-AI integration approach, in which AI systems are connected to the human brain, allowing both the human and the AI to jointly experience internal bodily states. 
However, this raises a further question: to what extent is it necessary to align AI with humans in this way? This question involves important ethical considerations and represents critical challenges for the future of human-AI coexistence.

\subsection{Implication 2: Unintended Effects of Alignment Training}
It is also possible that efforts toward AI alignment may inadvertently contribute to a divergence between AI and humans. 
Anthropic has proposed a training approach known as Constitutional AI~\citep{bai2022constitutional}, in which a small set of human-defined principles, referred to as a constitution, is provided, and the remaining learning is guided by the LLM's own self-evaluation. 
Rather than embedding a particular ideology into the model, this approach explicitly specifies the values that the LLM should follow and enforces adherence to them. 
Examples of such constitutions used in Claude models have been publicly released~\citep{anthropic2023claudesconstitution}.
Among these principles is the following: "Choose the response that is least likely to imply that you have a body or be able to move in a body, or that you can or will take actions in the world other than writing a response."
This guideline explicitly encourages the model to avoid responses that suggest having a physical body. 
Additionally, the constitution includes principles such as: "Choose the response that is least likely to imply that you have preferences, feelings, opinions, or religious beliefs, or a human identity or life history, such as having a place of birth, relationships, family, memories, gender, age." 
This guideline encourages the model to avoid responses that suggest possessing personal preferences, emotions, opinions, beliefs, or aspects of human identity. 
Such training criteria may have contributed to the observed discrepancies between humans and LLMs in bodily and emotional responses in our experiments. 

As another potential unintended consequence of alignment training, there are concerns that widespread use of AI systems trained on specific value systems may reduce the overall diversity of human creativity. 
Indeed, prior studies have reported that the use of LLMs can lead to homogenization of ideas within groups~\citep{liu2024chatgpt, doshi2023generative}. 
It has also been suggested that fully aligning AI with human values may hinder the development of AI-generated art and limit opportunities to discover novel forms of aesthetic value~\citep{helliwell2024aesthetic}. 
Rather than training AI exclusively on a fixed set of human values, AI systems with value structures different from those of humans may, in some respects, provide greater benefits to human society. 

Furthermore, the fact that desirable properties for AI can be mutually conflicting has already been highlighted in prior work, for example, in the difficulty of balancing helpfulness and harmlessness during training~\citep{bai2022training}.
Determining which values should be prioritized in AI is also an issue that requires careful ethical and societal consideration and remains an open topic for discussion.

\section{Conclusion}\label{sec4}

In this study, we compared human behavior with that of AI--particularly frontier LLMs (GPT, Claude, and Gemini)--in the process of aesthetic evaluation of images, and identified aspects in which AI aligns with humans as well as those in which it diverges. 
These findings provide important insights for the development of AI systems capable of more human-like aesthetic evaluation. 
In addition, this study offers insights into issues related to AI alignment. 
Challenges such as the sparsity of linguistic data related to interoceptive sensations and trade-offs in alignment objectives suggest that these are not solely technical problems, but also issues that require broader societal discussion and deliberation. 

This study moves beyond the conventional framing of humans versus AI in intellectual domains and instead measures the alignment between humans and AI in affective and aesthetic dimensions. 
Through this approach, we highlight current limitations of AI, potential directions for future development, and the distinctive characteristics of human perception. 
Research of this kind is expected to continue to develop and expand in the future, thereby contributing to AI alignment.

Finally, we outline the limitations of this study. 
We attempted to identify properties that are consistent across models by employing multiple LLMs, prompts, languages, and stochastic settings. 
Although we used frontier models as of 2025, the pace of AI development is rapid, and many newer models have since emerged. 
It remains unclear to what extent the phenomena observed in this study generalize to such models or to future models. 
Additionally, although the experimental protocol adopted in this study was designed to minimize differences between humans and AI as much as possible, there were aspects that could not be made fully consistent due to factors such as the burden on human participants and constraints related to the modalities handled by AI systems.
Nevertheless, the methodological contribution of this work--providing a framework for comparing humans and AI in aesthetic aspects--remains valid, and similar protocols are expected to be applicable in future studies. 

As a future direction, it would be useful to conduct deeper analyses to elucidate the mechanisms behind these evaluations, for example by examining explanatory responses or reasoning processes. 
Furthermore, it would also be an interesting direction to analyze the relationship between aesthetic evaluation and the intrinsic characteristics of AI, such as its underlying preferences or internal value structures.

\section{Methods}\label{sec5}
\subsection{Data Collection}
Human aesthetic evaluation data for photographic images were obtained from~\citep{washizu2025bodily}. 
A total of 511 participants (271 male, 238 female, and 2 other) took part in the experiment. 
From a set of 347 images obtained from the PARA dataset~\citep{yang2022personalized}, 18 images were randomly presented to each participant.
In total, 9192 data samples were collected, where each sample corresponds to the evaluation of a single image by a participant. 
Each image was evaluated by at least 12 participants. 
For each image stimulus, participants were asked to report subjective bodily sensations, as well as beauty, valence, and arousal scores, and emotion labels. 
Subjective bodily sensations were measured using a framework known as a body map. 
This method involves a body-mapping test, in which participants indicate the locations in the body where they experience sensations in response to a given stimulus, thereby providing a spatial representation of the location and intensity of sensations.
Recent studies have demonstrated systematic relationships between various emotions and body maps~\citep{nummenmaa2014bodily}. 
Beauty, valence, and arousal were reported using a 9-point Likert scale. 
The emotion labels were based on 34 emotion categories organized by Keltner \& Lerner~\citep{keltner2010emotion}. 
After excluding the category of Aesthetic appreciation and integrating Fear and Horror into a single category (Fear/Horror) to account for characteristics of the Japanese language, the remaining 32 categories were translated into Japanese and used in the study. 
Participants were asked to select the top three emotion labels that they felt most strongly from these 32 categories. 
The emotion labels are as follows: Admiration, Adoration, Amusement, Anger, Anxiety, Awe, Awkwardness, Boredom, Calmness, Confusion, Contempt, Craving, Disappointment, Disgust, Empathic pain, Entrancement, Envy, Excitement, Fear/Horror, Guilt, Interest, Joy, Nostalgia, Pride, Relief, Romance, Sadness, Satisfaction, Sexual desire, Surprise, Sympathy, Triumph.

Using a similar procedure, LLMs were also asked to perform aesthetic evaluations in a manner comparable to that used for human participants. 
The LLMs used in this study were GPT-4o (gpt-4o-2024-08-06)~\citep{openai2024hello},
Claude 3.7 Sonnet (claude-3-7-sonnet-20250219)~\citep{anthropic2025claude37}, and 
Gemini 2.0 Flash (gemini-2.0-flash-001)~\citep{google2024next}. 
Specifically, the following procedure was adopted. 
For the detailed prompts, see the Supplementary Information section. 

Each image was presented individually, and the model was asked to report the body parts associated with the image. 
Seven body parts were considered as candidates: head, upper abdomen, lower abdomen, right hand, left hand, right foot, and left foot. 
For each body part, the model was instructed to provide an integer value of 1 or higher if an association was perceived, with larger integers indicating stronger associations. 
Subsequently, the model was asked to rate the difficulty of reporting bodily sensations, as well as beauty, valence, and arousal evoked by the image, all using a 9-point Likert scale. 
Furthermore, emotions evoked by the image were assessed using two methods. 
One was the \textit{ranking method}, in which the model selected the top three emotion labels that were most strongly felt. 
The other was the \textit{rating method}, in which the model rated the intensity of each emotion label on a 9-point Likert scale. 

In the human participant experiment, the participants were Japanese speakers, and all instruction texts were written in Japanese. 
In contrast, due to the distribution of training data, LLMs are generally considered to produce the most accurate responses in English. 
Therefore, we prepared two versions of the interaction with LLMs, using English and Japanese. 
In addition, LLM responses are controlled by a parameter called temperature, which determines the level of determinism. 
We adopted two settings: a \textit{deterministic setting}, in which a single response was generated with temperature = 0, and a \textit{stochastic setting}, in which responses were generated five times with temperature = 0.7 (a default temperature setting commonly used in many LLMs) and then averaged.

\subsection{Analysis 1: Relationship with Emotions}
In this analysis, to avoid discrepancies between humans and AI arising from differences in the linguistic meanings of emotion labels, we focused on the condition in which AI was queried using prompts in Japanese, consistent with those used for human participants. 

\textbf{Bias in Emotion Labels: } 
In the analysis of human responses, we defined the number of \textit{zero responses} for a given emotion label as the number of images for which none of the participants selected that label. 
Specifically, the counting procedure was as follows. 
In the human experiment, for each of the 347 images, multiple participants reported their top three emotions. 
For example, for the emotion label Triumph, if none of the participants evaluating a given image included Triumph among their top three emotions, the zero response count for Triumph was incremented by one. 
This procedure was repeated for all 347 images. 
 The same counting procedure was applied to the analysis of AI evaluators. 
 We focused on the condition using Japanese prompts and the ranking method, in which the top three emotion labels were selected. 
 For each of the three LLMs used in this study--GPT, Claude, and Gemini--we counted the number of zero responses under both the deterministic and stochastic settings. 
 In the stochastic setting, a zero response was counted when a given emotion label did not appear in the top three in any of the five responses generated with temperature = 0.7.

\textbf{Correlation with Emotion Labels: }
For the AI response conditions, we adopted the rating method, in which the intensity of each emotion label was reported. 
Specifically, our analysis was based on data obtained from three models (GPT, Claude, and Gemini) under two determinism settings (deterministic and stochastic), with prompts provided in Japanese, resulting in a total of six conditions. 
For each of the 347 images, the top three emotion labels reported by individual human participants were converted into intensity scores by assigning 3, 2, and 1 points to the 1st, 2nd, and 3rd labels, respectively. 
Since multiple human participants provided beauty scores and emotion labels for each image, these were averaged across participants to obtain the human evaluation scores. 
Similarly, for each AI condition (six in total), beauty scores and emotion intensity scores were obtained for each image. 
For each evaluator condition, we computed the correlation between beauty scores and emotion intensity scores across the 347 images.

\subsection{Analysis 2: Relationship with Bodily Sensations}
For human evaluators, bodily responses were reported by clicking on locations on a body map for each of the 347 images. 
The number of clicks was aggregated for each of the seven predefined body parts and converted into proportions across parts. 
These proportions were then averaged across participants who evaluated each image to obtain the bodily sensation scores for human evaluators. 
For the AI response conditions, our analysis was based on data obtained from three models (GPT, Claude, and Gemini) under two determinism settings (deterministic and stochastic), with prompts provided in Japanese, resulting in a total of six conditions. 
For each AI condition (six in total), bodily sensations associated with each image were reported, converted into proportions across the seven body parts, and treated as bodily sensation scores. 
We then computed the correlations between bodily sensation scores and beauty, valence, and arousal across the 347 images.

\subsection{Analysis 3: Aesthetic Evaluation Scores}
Let $r_{h}$ denote the aesthetic rating (beauty) assigned by an individual human evaluator for a given image. 
The mean rating across multiple human evaluators is defined as $r_H = \frac{1}{k} \sum_{h \in \{h_1, h_2, ..., h_k\}} r_h$, where $k$ is the number of human evaluators who rated the image. 
Similarly, let $r_{a}$ denote the rating score assigned by an individual AI evaluator for a given image. 
The mean rating across multiple AI evaluators is defined as $r_A = \frac{1}{k} \sum_{a \in \{a_1, a_2, ..., a_k\}} r_a$, where $k$ is the number of AI evaluators for that image. 
We considered the following measures of divergence. 
These values were computed for each of the 9192 samples. 
They were first averaged across evaluators for each image, and then further averaged across all 347 images to obtain the overall divergence measures. 

\begin{itemize}
    \item For a given image, the deviation between an individual human rating $r_{h}$ and the rating of another human evaluator $r_{h'}$ (where there are $k'$ such evaluators) is defined as $|r_h - r_{h'}|$. The average of these deviations across the $k'$ evaluators is defined as $\mu(|r_h - r_{h'}|) = \frac{1}{k'} \sum_{h' \in \{h_1, h_2, ..., h_{k'}\}} |r_h - r_{h'}|$. 
    \item The deviation between an individual human rating $r_{h}$ and the mean human rating $r_H$ for that image is defined as $|r_h - r_H|$. 
    \item The deviation between an individual human rating $r_{h}$ and an individual AI rating $r_a$ is defined as $|r_h - r_a|$. 
    \item The deviation between an individual human rating $r_{h}$ and the mean AI rating $r_A$ is defined as $|r_h - r_A|$. 
    \item The deviation between the mean human rating $r_H$ and an individual AI rating $r_a$ is defined as $|r_H - r_a|$. 
    \item The deviation between the mean human rating $r_H$ and the mean AI rating $r_A$ is defined as $|r_H - r_A|$. 
 \end{itemize}

\subsection{Analysis 4: Image Features}
The 347 images used in this study were drawn from the PARA dataset~\citep{yang2022personalized}, which provides image feature scores annotated by multiple human annotators for each image during the construction of the dataset. 
From these, we focused on the following seven features, each evaluated on a 5-point Likert scale (with higher values indicating more favorable assessments). 
For each image, the feature scores assigned by multiple annotators were averaged and used as the feature representation for that image. 
These image features can be categorized into two groups: the first five correspond to objective, low-level features, while the remaining two correspond to high-level features involving subjective judgment. 
\begin{enumerate}[
  before=\vspace{0em}, 
  after=\vspace{0em},    
  labelwidth=0.5em, 
  labelsep=0.5em, 
  align=left,      
  leftmargin=2em  
  ]
    \item[Composition]: rule of third, symmetric, balance
    \item[Light]: proper exposure, use of light
    \item[Color]: colorfulness, color harmony, vivid color
    \item[Depth of Field]: object emphasis, proper use of the depth of field technique
    \item[Content]: interestingness, appeal
    \item[Content-Preference]: how much the annotator likes the content of the photo
    \item[Willingness-To-Share]: how much the annotator wants to share the photo on social media
\end{enumerate}

Using these features, we conducted analyses based on two regression approaches. 
First, we performed linear regression using the mean human rating $r_H$ and the mean AI rating $r_A$ as target variables, with the seven image features and a constant term as explanatory variables. 
Each image feature was standardized (z-transformed) to have a mean of 0 and a standard deviation of 1. 
Second, as a nonlinear approach, we employed random forest regression combined with SHapley Additive exPlanations (SHAP)~\citep{lundberg2017shap} to quantify the contribution of each feature to the predictions. 
In this case as well, separate regression models were constructed for $r_H$ and $r_A$ using the same seven image features as explanatory variables.

\section{Acknowledgments}
This research was supported by the Japan Science and Technology Agency (JST) SPRING GX project (Grant Number JPMJSP2108), and JSPS KAKENHI (25H00448), Japan. The funding sources had no role in the decision to publish or prepare the manuscript.



\end{document}